\documentclass[%
 aip,
 jmp,%
 amsmath,amssymb,
 reprint,%
]{revtex4-2}

\usepackage{graphicx}
\usepackage{dcolumn}
\usepackage{bm}
\usepackage{comment}
\usepackage{float}
\usepackage[section]{placeins}

\begin{document}


\title{Photo-induced molecular growth of benzonitrile in the gas phase}

%

\author{Nihar Ranjan Behera}%
\altaffiliation{%
 Indian Institute of Technology, Madras\\
 Chennai 600036, India 
}
\author{Arun Kumar Kanakati}%
\altaffiliation{%
 School of Chemistry, University of Hyderabad\\
 Hyderabad 500046, India 
}
\author{Pratikkumar Thakkar}%
\altaffiliation{%
 Indian Institute of Technology, Madras\\
 Chennai 600036, India 
}
\author{Siddhartha Sankar Payra}%
\altaffiliation{%
 Indian Institute of Technology, Madras\\
 Chennai 600036, India 
}
\author{Saurav Dutta}%
\altaffiliation{%
 Indian Institute of Technology, Madras\\
 Chennai 600036, India 
}
\author{Saroj Barik}%
\altaffiliation{%
 Indian Institute of Technology, Madras\\
 Chennai 600036, India 
}
\author{Yash Lenka}%
\altaffiliation{%
 Indian Institute of Technology, Madras\\
 Chennai 600036, India 
}
\author{G Aravind}%
\altaffiliation{%
 Indian Institute of Technology, Madras\\
 Chennai 600036, India 
}

\date{\today}

\begin{abstract}
Due to the absorption of high energetic ultraviolet (UV) photons by the surface layers of the cold molecular clouds, only low energetic photons are able to penetrate into the inner regions of these clouds. This leads to lower photo-ionization yield of molecules of higher ionization potential in these environments. However, here we have experimentally shown the ionization of Benzonitrile molecule using 266nm (4.66eV) photons. The low intensity and unfocused laser irradiation of benzonitrile molecules results extensive fragmentation. Moreover, the ion-neutral reactions among the cationic fragments and neutral fragments shows promising molecular mass growth.
\end{abstract}

\maketitle

\section{Introduction}\label{sec:intro}
Polycyclic aromatic hydrocarbons (PAHs) are the missing link between the small carbon molecules, fullerenes, and the carbonaceous nanoparticles (interstellar grains), as postulated by the Puget and L\'eger for the very first time in the year 1989 \citep{puget1989new}. PAHs are considered to be abundant in the interstellar medium (ISM), and are believed to make up $\sim 25 \%$ of the cosmic carbon of our galaxy \cite{dwek1997detection}. The 3-20$\mu m$ unidentified infrared emission bands (UIBs) of the ISM are more often attributed to the PAHs \cite{allamandola1989interstellar}, \cite{peeters2011pah} including their derivatives and ionic forms, due to the close match up of the characteristic vibrational frequencies of C-C and C-H bonds of the aromatics with the UIBs. PAHs are also the potential carriers of the diffuse interstellar bands \citep{salama1999polycyclic}-discrete absorption features superimposed on the interstellar extinction curve. The presence of PAHs in many carbonaceous chondrites is a strong evidence of their origin in circumstellar environments \citep{parker2017formation,elsila2005alkylation,spencer2008laser,zenobi1989spatially}. Primarily, the synthesis of PAHs occur at elevated temperatures of few 1000K, and dense circumstellar envelope of asymptotic giant branch (AGB) stars, and at planetary nebulae \citep{tielens2008interstellar,latter1991large,joblin2011formation,frenklach1989formation,cherchneff1992polycyclic}, through hydrogen abstraction carbon addition mechanism (HACA)\citep{frenklach2002reaction}. However, PAHs have lifetimes of few orders of $10^8$ years \citep{micelotta2010polycyclic,joblin2011formation}, as they are destroyed by galactic shock waves, cosmic rays, and photolysis. These lifetimes are considerably shorter as compared to the rate of injection of PAHs from circumstellar envelopes of carbon-rich AGB stars \citep{joblin2011formation}, \textit{i.e.} the time scales of $2\times10^9$ years. Thus, the omnipresence of PAHs in the ISM refers to hitherto unknown pathways of formation and growth of PAHs in those environments. The same is inferred from the recent observation of higher derived column densities of benzonitrile \citep{mcguire2018detection} and 1- and 2-Cyanonaphthalene (CNN) \citep{mcguire2021detection} as opposed to the current astrochemical models. So, it is crucial to identify the routes of formation of PAHs to understand their origin in ISM. There is a widely accepted two step mechanism associated with the growth of PAHs \citep{cherchneff1992polycyclic} at low temperatures. The first step involves the closure of the first aromatic ring, producing either benzene (C\textsubscript{6}H\textsubscript{6}) or phenyl (C\textsubscript{6}H\textsubscript{5}) radical in general. In a subsequent step, the closure of the second aromatic ring and further growth of PAHs are considered. These reactions essentially involve neutral reacting partners. However, molecules undergo fragmentation and ionization processes that form both neutral and ionic species. So, it is very much expected that reactions involving these species might account for the growth of PAHs in ISM.

First detected in the Taurus molecular cloud (TMC-1) \citep{mcguire2018detection}, benzonitrile is also the first aromatic molecule detected using radio astronomy. Later, it was also detected in other molecular clouds namely Serpens 1A, Serpens 1B, Serpens 2 and MC27/L1521F \citep{burkhardt2021ubiquitous}. Benzonitrile is suggested to be formed via a barrierless reaction between CN radical and C\textsubscript{6}H\textsubscript{6} \citep{balucani1999crossed,trevitt2010reactions,mcguire2018detection,lee2019gas,cooke2020benzonitrile}, which is why, it is regarded as the proxy for the radio-inactive benzene molecule. In this way benzonitrile linked to the fundamental aromatic ring \textit{i.e.} benzene molecule could provide observational constraints at the early stages of aromatic chemistry in the ISM. Including benzonitrile, there are around 92 molecules detected towards and in TMC-1\footnote{http://hutchinson.belmont.ma.us/tth/}. Among them, molecules like cyanoacetylene (HNCCC) \citep{kawaguchi1992detection}, 1- and 2-cyano-cyclopentadiene are of particular interest as we explore the chemical network surrounding benzonitrile. Here, we investigated possible routes of formation of above mentioned molecules through photo-fragmentaion of benzonitrile and ion-neutral reactions involving photo-fragments.

\section{Experimental methods}\label{Ch5.2Exp_methods}

\begin{figure}[H]
	\centering
	\includegraphics[width=1\linewidth]{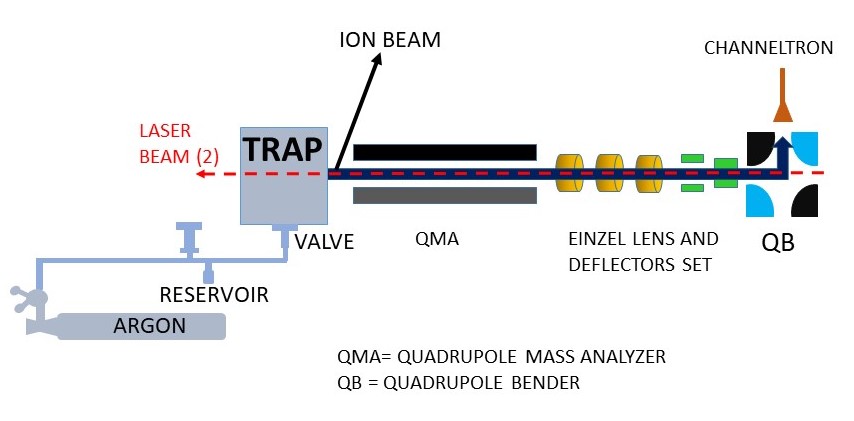}
	\caption{Schematic of experimental setup employed for ion-neutral reactions involving benzonitrile and its photo-fragments at stage 1.}
	\label{fig:schematic_fig5.1}
\end{figure}

The newly built 22-pole rf ion trap experimental setup was employed for the present studies on ion-neutral reactions between benzonitrile and its photo-fragments. Detailed descriptions of the experimental setup can be found in our previous work \citep{behera202422}. Only necessary descriptions of the present experiments are presented here. The experiments were conducted in two stages. Stage 1: A reservoir containing high-purity ($99\%$) benzonitrile was heated to its boiling point to generate vapour. Subsequently, the vapour of the benzonitrile was seeded with \textit{Argon} carrier gas such that the stagnation pressure of the mixture was maintained between $0.25-1.5$ $Kg.cm^{-2}$. The mixture was introduced into a 22-pole rf ion trap through a general pulsed valve running at a 10Hz repetition rate, as shown in figure \ref{fig:schematic_fig5.1}. The valve employs a flat nozzle with an aperture diameter of 0.8mm, and alongside the lower stagnation pressure maintained behind the nozzle, avoids clustering. An unfocused $266$ $nm$ laser light (laser beam (2) in figure \ref{fig:schematic_fig5.1}) passing through the ion trap along its axis, illuminated the gas mixture inside it. The cationic fragments generated from neutral-photon interactions were swiftly extracted towards the detector by applying $+100$ $V$ to the entrance electrode and 2MHz rf signal with a DC offset of $+95$ $V$ to 22-poles of the ion trap. Prior to detection, the ions were mass-analyzed using a quadrupole mass analyzer (QMA)

Later, ions were trapped over a few milliseconds, allowing reactions between cations and neutrals produced due to the photoexcitation of benzonitrile. The products of ion-neutral reactions were mass analyzed by QMA. The yield of each product was monitored with increasing storage time to get an overall idea of chemical evolution.

\begin{figure}[H]
	\centering
	\includegraphics[width=1\linewidth]{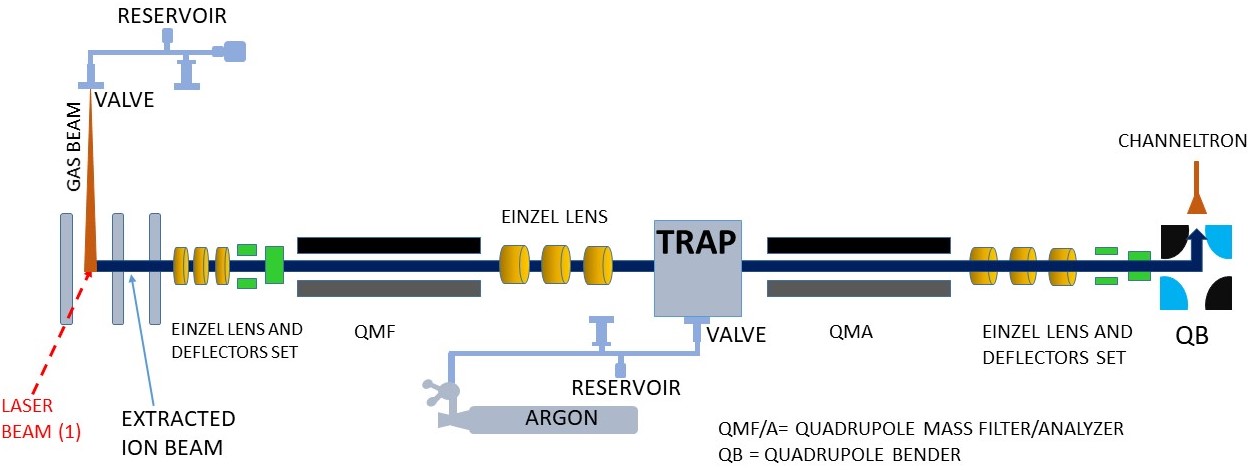}
	\caption{Schematic of experimental setup employed for ion-neutral reactions involving benzonitrile and its photo-fragments at stage 2.}
	\label{fig:schematic_fig5.2}
\end{figure}

Stage 2: While the experiments in Stage 1 allowed reactions between photo-fragments, the direct relationship between specific reactants and products was unclear. To investigate the direct relationship, ions were generated outside the ion trap, and a particular ion (e.g., protonated parent cation) was introduced into the ion trap. The ions were produced in the same manner as in Stage 1, with an unfocused $266$ $nm$ laser beam intersecting with the pure benzonitrile vapour (see figure \ref{fig:schematic_fig5.2}). The ion of interest (protonated parent cation) was filtered and loaded into the ion trap allowing to react with benzonitrile neutral carried into the trap by Argon, similar to Stage 1. Mass analysis of the products was performed using the QMA, and the yields of each product were monitored over time. Throughout the experiments, the laser intensity was varied within a range of $1 \times 10^{5}-6 \times 10^{6}$ $W.cm^{-2}$.

\section{Results and Discussions}\label{Ch5.3Results_and_discussions}

In this experiment, resonant $\pi - \pi ^*$ excitation of  benzonitrile neutrals to their S$_1$ states \citep{rajasekhar2022spectroscopic} led to the formation of different cationic and neutral fragments. Mass spectroscopic analysis revealed the formation of cations at m/z of $39u$, $50u$ $51u$, $52u$, $66u$, $76u$, $104u$, and $127u$, while we did not detect the neutrals in our experiment. First, we discuss the formation of these fragments. The ionization energy of benzonitrile neutral is $\sim 9.72$ $eV$ \citep{araki1996two, kamer2023threshold}, indicating a requirement of three photons for its ionization.  Additionally, the appearance energies of fragments smaller than the parent are found to be high, around $\sim$ $13.8-17$ $eV$ \citep{kamer2023threshold}, implying a need for four photons for their formation under the experimental conditions. However, we employed unfocused laser pulses with intensities varying between $10^{5}-10^{6}$ $W cm^{-2}$, and given the three and four photons requirement for the observed cations, MPI or REMPI are unlikely to be operational. Furthermore, the formation of $127u$ cation (heavier than parent) can not be explained via REMPI.

\begin{figure}[H]
	\centering
	\includegraphics[width=1\linewidth]{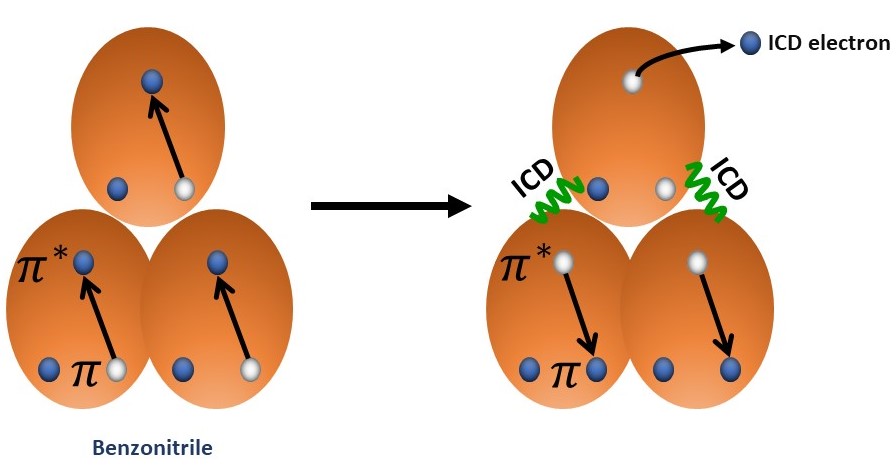}
	\caption{Schematic of three-centre ICD mechanism in which two excited benzonitrile monomers decay by transferring energy to a third neighbour excited monomer and ionizing it.}
	\label{fig:sicd_mechanism_fig5.3}
\end{figure}

So, we now discuss the role of intermolecular coulombic decay (ICD) in photoexcited benzonitrile monomers, which resulted in the formation of different cations in this experiment. ICD is a non-local electronic decay process wherein an inner-valence vacancy is filled by an outer-shell electron and the excess energy ionizes the neighbor atom or molecule \citep{cederbaum1997giant}. The $\pi - \pi^*$ excitation of benzonitrile monomers induce long range covalent interactions among the excited monomers. This leads to the association of three excited monomers, and \textit{en route} to the association, one or two of the excited units decay by transferring energy via ICD to the neighbor unit and eventually ionizing the trimer. The trimer cation is unstable and fragmented into different fragments including $104u$ and $127u$ cations. $104u$ cations are formed as a result of hydrogen migration within the trimer following the fragmentation.  This channel of ICD in unbound molecules such as pyridine \citep{barik2022ambient} and quinoline \citep{barik2023molecular}, has already been reported in our previous studies where covalent interactions among photoexcited monomers resulted in efficient molecular mass growth. ICD via concerted transfer of electronic energy from multiple photoexcited benzonitrile monomers to a ground state benzonitrile molecule \citep{barik2023intermolecular} leads to dissociative ionization of the molecule. These two ICD channels compete with each other, and at low laser intensity the ICD channel via concerted energy transfer, overtakes the former ICD channel \citep{barik2023intermolecular}. ICD occurs at sub-femtosecond timescale and is a long-range collisionless phenomenon. So, the hard-sphere collision model does not apply here for ionization to occur in the present experiments.

Now, the cations are allowed to react with the neutrals of benzonitrile and its photo-fragments as we stored them inside a 22-pole rf ion trap for a few milliseconds. Before we discuss the results of the experiment, it is to be noted that cations with m/z below $66u$ did not participate in ion-neutral reactions here. This is because those ions were not stored inside the ion trap as we applied rf voltages of 2 MHz frequency to the 22-poles of the ion trap. This was evident from the observation that almost all those ions were lost over a few microseconds of storage time.

\begin{figure}[H]
	\centering
	\centering
	\includegraphics[width=0.9\linewidth]{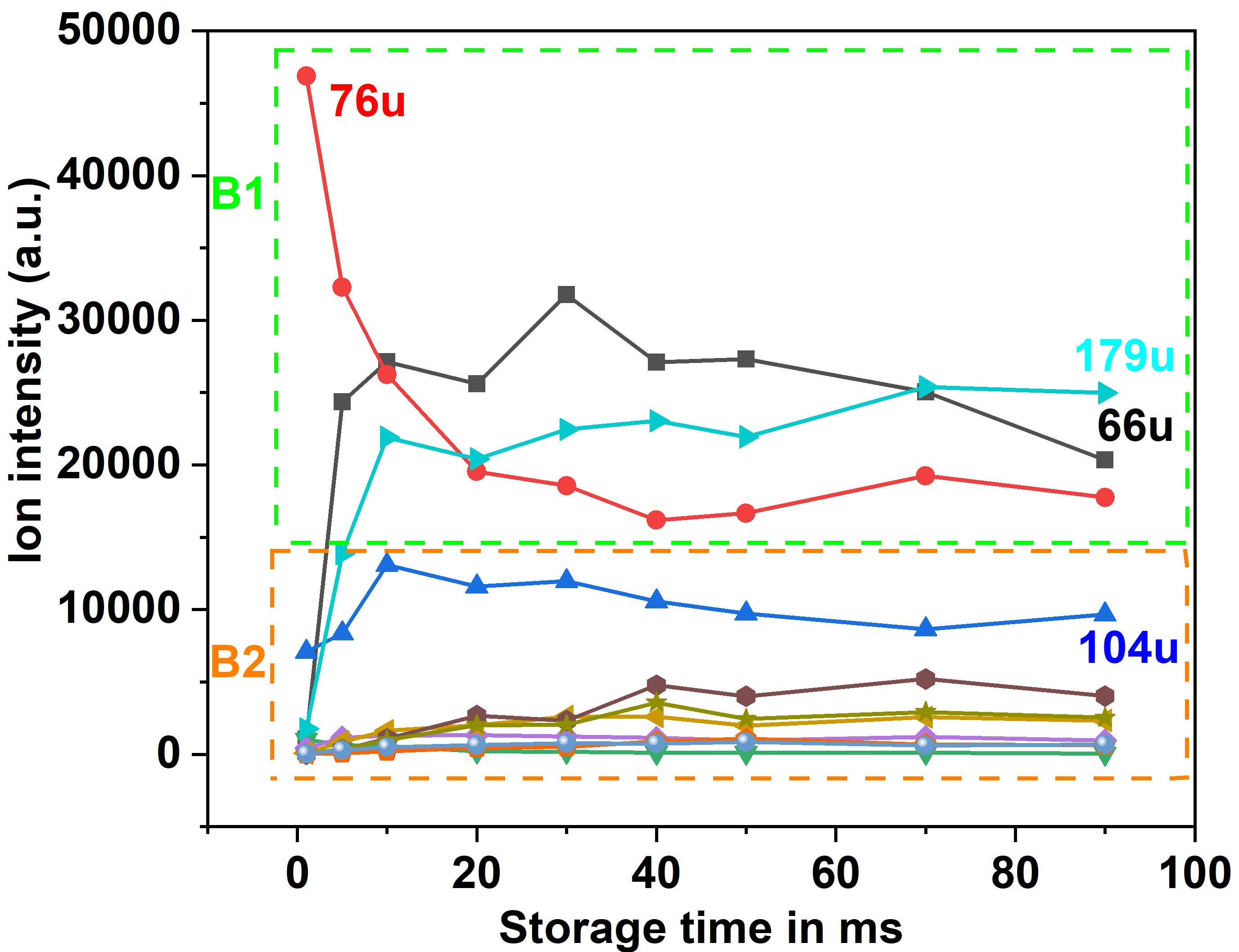}
	\caption{Temporal evolution of reactant and product ions of ion-neutral reactions involving benzonitrile and its photo-fragments.}
	\label{fig:benzonitrile_trapping_1}
\end{figure}

We first concentrate our discussions on the green box (B1) of figure \ref{fig:benzonitrile_trapping_1}, which shows the rapid decrease of $76u$ cations and increase of $66u$ and $179u$ cations over a few milliseconds of storage time. Based on these observations, we propose the following reaction pathways. The m/z $76u$ cation (HCN loss channel) is a dominant cation formed as a result of dissociative ionization of benzonitrile neutral via ICD. Bicyclic meta-benzyne$^{.+}$ (BMB) and meta-benzyne$^{.+}$ are two isomers that could be formed with BMB cation energetically more favorable \citep{kamer2023threshold}. So, the $179u$ cations are formed through the collision between BMB cations and benzonitrile neutrals (see reaction \ref{eq_5.1}). Zhen et. al.\cite{zhen2019laboratory} has observed the formation of such cluster cations through ion-neutral reactions between dicoronylene cations and anthracene neutrals.

\begin{equation}\label{eq_5.1}
	\centering
	C_6H_4^+ + C_7H_5N \rightarrow C_{10}H_9N^+
\end{equation}

Now, we consider the evolution of the $66u$ cations. The $C_3H$ loss channel leads to the formation of $C_4H_4N^+$ cation. This channel was not reported in earlier studies on dissociative photoionization of benzonitrile \citep{kamer2023threshold} and a very low yield of the $66u$ was observed in the case of electron impact ionization \citep{rap2023fingerprinting}. The rapid increase in the yield of the $66u$ cations, exceeding the yield of all other ions except the $76u$ cations, clearly indicates participation of the  $76u$ cations in the formation of 66u cations. Here, we put forward a bimolecular reaction between 76u cations and benzonitrile neutrals to form the $66u$ cations.

\begin{equation}\label{eq_5.2}
	\centering
	C_6H_4^+ + C_7H_5N \rightarrow 66u^+ + 113u
\end{equation}

Two possible isomeric product channels account for the reaction \ref{eq_5.2} are $C_5H_6^+ + C_8H_3N$ and $C_4H_4N^+ + C_9H_5$. $C_9H_5$ radical has visible absorption bands that coincide with the weak diffuse interstellar bands toward HD183143 and HD204827 \citep{steglich2016visible}. Similarly, $C_5H_6^+$ and $C_4H_4N^+$ are two potential stoichiometries which could contribute to the m/z signal at $66u$ in Titan's atmosphere, measured using ion-neutral mass spectrometer in the recent  Cassini-Hyugens mission \citep{hendrix2020cation}. The rapid increase of product ions of the reactions between BMB cations and benzonitrile neutral points toward higher kinetic rates of the reactions. The time evolution of both the product ions also reveals that they are highly competing product channels.

\begin{figure}[H]
	\centering
	\centering
	\includegraphics[width=1\linewidth]{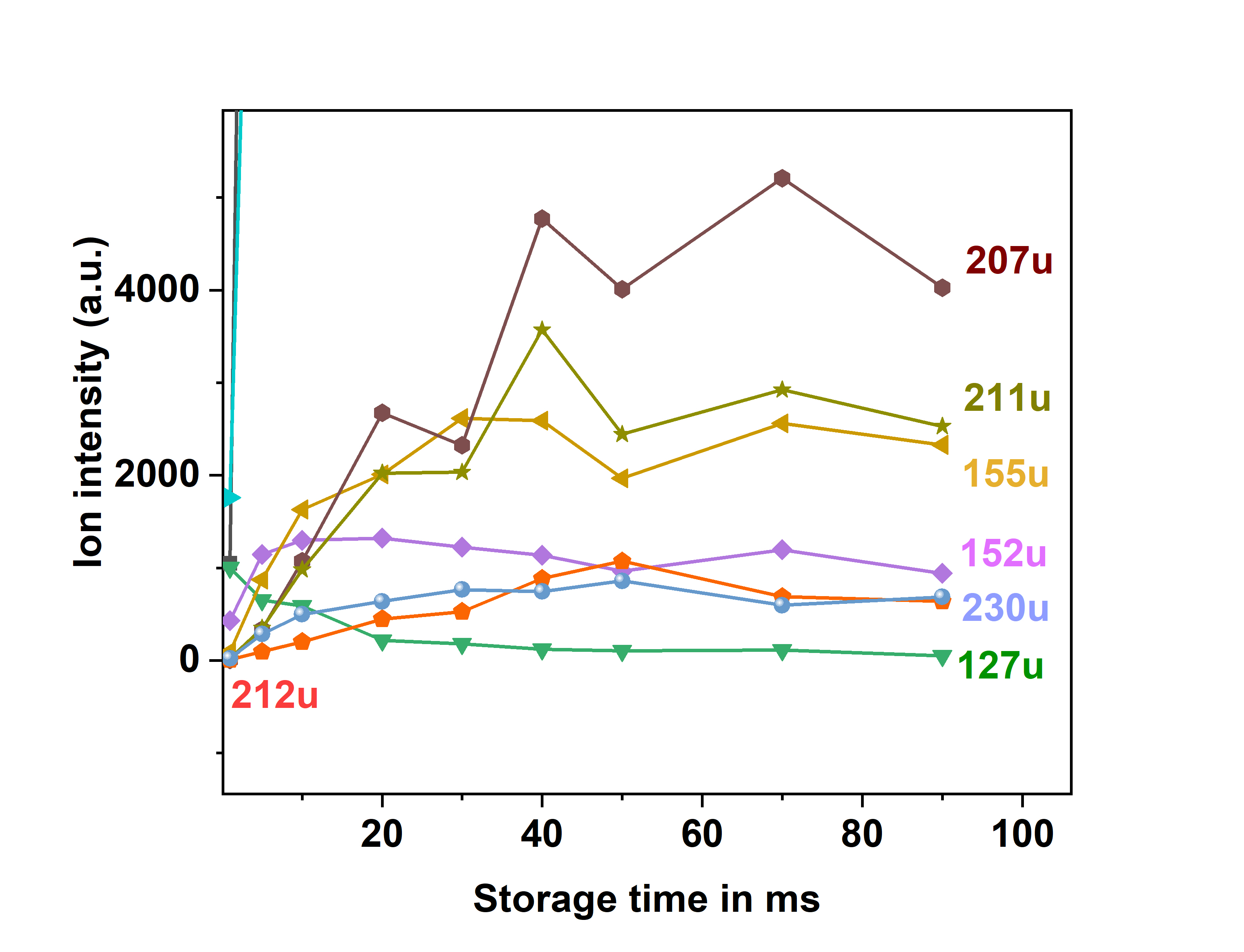}
	\caption{Temporal evolution of reactant and product ions of ion-neutral reactions involving benzonitrile and its photo-fragments.}
	\label{fig:benzonitrile_trapping_2}
\end{figure}

Now that we have discussed the formation routes of two dominant product ions ($66u$, $179u$), we now examine the temporal evolution of ions as specified by the orange box (B2) of figure \ref{fig:benzonitrile_trapping_1} (also see figure \ref{fig:benzonitrile_trapping_2}).
 The ions were prepared outside the ion trap (see figure \ref{fig:schematic_fig5.2}), and the mass selected 104u cations were loaded into the trap. The $104u$ cations were allowed to react with benzonitrile neutrals, which were introduced into the trap through Argon carrier gas.  The reaction resulted in the formation of the $207u$ cations, but we did not observe the $211u$ and $212u$ cations. Thus, a series of bimolecular reactions between the $207u$ cations and hydrogen (molecular or atomic) could lead to the formation of the $211u$ and $212u$ cations in our experiment, as aromatic cations readily react with  hydrogen (molecular or atomic) \citep{snow1998interstellar}. Typically, Hydrogen (molecular and atomic) loss channels are few eV above the ground state of PAH \citep{simon2018energetic,trinquier2017pah,trinquier2017pah2}.

Likewise, the bimolecular reactions between the $127u$ cations and benzonitrile neutrals lead to the formation of the $230u$ cluster ions. In the end, we investigate the formation of $152u$ and $155u$ cations in the present experiment. The yield of $152u$ and $155u$ cations going past the yield of the $127u$ cations points out the non-involvement of the $127u$ cations in the formation of 152u and 155u cations. Based on these observations, we propose the reactions between 104u cations and neutral fragments corresponding to $50u$, $51u$, and $52u$ cations for forming $152u$ and $155u$ cations. Their yield is low and shows a slow increase over storage time. So, the reactions have the slowest kinetic rates among all the reactions.

\section{Astrophysical implications}\label{Ch5.4astrophysical_implications}

Having discussed the formation of different cations through dissociative ionization and ion-neutral reactions, we now examine their astrophysical implications. As discussed earlier, the $66u$ cations are formed in the dissociative ionization of benzonitrile neutral through the $C_3H$ loss channel.  The $C_3H$ neutral and its cationic counterparts are detected in many astrophysical environments \citep{thaddeus1985astronomical,yamamoto1987laboratory,pety2012iram,brunken2014laboratory,cernicharo2022discovery}. The fact that both $C_3H$ (linear and cyclic form) and benzonitrile are detected in the same TMC-1, our observations clearly show that dissociative photoionization of benzonitrile is undoubtedly one of the routes to form  $C_3H$ radical in TMC-1. 

The $152u$ and $155u$ cations, whose masses are close to the mass of 1- and 2-CNN (153u), may be dehydrogenated and 2H additive counterparts of the CNN molecule. Similarly, the $52u$ cation observed in the photoionization of benzonitrile in the present experiment may account for protonated cyanoacetylene $HC_3NH^+$ (52u) which is detected in TMC-1 \citep{kawaguchi1994detection}. Although we do not provide IR spectra of these molecules, which could confirm our predictions, they are potential candidates for the detection in TMC-1.

\section{Conclusions}\label{Ch5.5conclusions}

In conclusion, dissociative photoionization of benzonitrile using 266nm photons via ICD was observed. The study involved unfocused low intensity laser pulses. Additionally, we studied the ion-neutral reactions between benzonitrile and its phto-fragments. The importance of product ions formed in the present experiments that could be accounted for TMC-1 and different formation routes were also discussed. Hopefully, our observations will trigger theoretical and experimental investigations that can confirm all the possible routes discussed in this chapter.

\end{document}